\documentclass[preprint,12pt,a4paper,byrevtex]{revtex4}%
\usepackage{natbib}
\usepackage{graphicx}
\usepackage{amsfonts}
\usepackage{amsmath}
\usepackage{amssymb}%
\providecommand{\U}[1]{\protect\rule{.1in}{.1in}}
\begin{document}
\preprint{ }
\title[ ]{Tunneling in double barrier junctions with 'hot spots'}
\author{D. Herranz, F. G. Aliev}
\affiliation{Departamento F\'{\i}sica Materia Condensada, Universidad Aut\'{o}noma de
Madrid, 28049,\ Madrid, Spain }
\author{C. Tiusan, M. Hehn}
\affiliation{Institut Jean Lamour, Nancy Universit\'{e}, 54506 Vandoeuvre-l\`{e}s-Nancy
Cedex, France}
\author{V.~K.~Dugaev}
\affiliation{Department of Physics, Rzesz\'{o}w University of Technology, 35-959
Rzesz\'{o}w, Poland}
\author{J. Barna\'s}
\affiliation{Department of Physics, Adam Mickiewicz University 61-614 Poznan; and Institute
of Molecular Physics, 60-179, Poznan, Poland}
\keywords{Double barrier magnetic tunnel junctions;\ Quantum well states, hot spots}
\pacs{PACS number: 73.21.-b; 73.63.Rt;\ 72.25.Mk}

\begin{abstract}
We investigate electronic transport in epitaxial Fe(100)/MgO/Fe/MgO/Fe double
magnetic tunnel junctions with soft barrier breakdown (hot spots). Specificity
of these junctions are \emph{continious middle layer }and Nitrogen doping of
the MgO\ barriers\textbf{ }which provides soft breakdown at biases about 0.5V.
In the junctions with hot spots we observe quasi-periodic changes in the
resistance as a function of bias voltage which point out formation of quantum
well states in the middle Fe continuous free layer. The room-temperature
oscillations have been observed in both parallel and antiparallel magnetic
configurations and for both bias polarizations.\textbf{ } A simple model of
tunneling through hot spots in the double barrier magnetic junction is
proposed to explain qualitatively this effect.

\end{abstract}

\maketitle

Recent theoretical predictions \cite{Butler,Mathon} followed by experimental
observations of coherent tunneling in magnetic tunnel junctions (MTJs) with
MgO barriers \cite{Bowen,Faure-Vincent,Parkin,Yuasa,GuerreroAPL}\textbf{ }
have busted research of $F_{1}/I/F_{2}$ MTJs \cite{Moodera} (here $F_{1}$ and
$F_{2}$ are ferromagnetic layers and $I$ - insulating barrier), and opened new
perspectives of their applications in spintronic devices. For thin enough
electrodes, electron tunneling may reveal resonant features due to quantum
well states (QWS). Early studies explored the simplest way to realize resonant
tunneling by growing a thin nonmagnetic layer (NM)\ between ferromagnetic
electrode and barrier in standard MTJ structures $F_{1}/NM/I/F_{2}$
\cite{LeClair,YuasaScience,Greullet}. However, resonant tunneling in double
MTJs (DMTJs) $F_{1}/I/F_{c}/I/F_{2}$ ($F_{c}$ is central layer) may have
advantages in comparison with the standard MTJs, mainly due to their enhanced
tunnelling magnetoresistance (TMR) \cite{Barnas,Petukhov,Kalitsov} and
resonant\ spin-torque effects \cite{VedyaevA,Theodonis,Watanabe}. TMR in DMTJs
only weakly varies at low bias voltages \cite{PeraltaRamos}, which is crucial
for applications. Last but not least, the current driven magnetization
reversal in DMTJs occurs at relatively low current densities \cite{Watanabe}.

In DMTJs the QWS can strongly influence electron transport only if the
\emph{F}$_{c}$ layer has submonolayer roughness and its thickness exceeds 1nm
minimizing Coulomb blockade (CB)\ effects \cite{G.Feng}. On the other hand,
the \emph{F}$_{c}$ layer should also be thin enough so that energy separation
of QWS substantially exceeds the thermal energy. These conditions are hardly
fulfilled in the macroscopic DMTJs \cite{Tiusan-JCM,Coey}, where resonant
tunneling was not observed mainly due to the absence of the atomically flat
surfaces over entire junction lateral dimensions. Evidence for a local
tunneling through QWS in the central Fe layer in $F_{1}/I/F_{c}/I/NM$
junctions was provided by Iovan et al. \cite{Iovan} using the point contact
technique.\textbf{ }Recently, Nozaki et al. \cite{Nozaki-PRL} reported on
resonant tunneling effects in macroscopic\textbf{ }DMTJs with Fe nano islands
incorporated into the thick MgO barrier. Their \ results have been interpreted
as due to combined QWS \cite{YanWang} and CB effects.

This letter reports on the detailed study of electron transport in epitaxial
macroscopic Fe$(100)$/MgO/Fe/MgO/Fe junctions with \emph{continuous middle Fe
layer }and current flowing through the 'hot spots'. Oscillatory conductance
and TMR with applied voltage present clear signatures of the local coherent
tunneling through QWS. The oscillations have been observed in \emph{both
parallel and antiparallel magnetic configurations and for both bias
polarizations}, contrary to Ref.\cite{Nozaki-PRL} where they were seen only in
the parallel configuration and for one current orientation. Moreover, we
observe oscillations in the room temperature (RT) regime, while in
Ref.\cite{Nozaki-PRL} they were seen mainly at low temperatures. Finally, the
technique we used for their observation is also different and makes use of
some features of breakdown junctions, which allow to observe \emph{quantum
oscillations in a continuous layer}. Our results have been explained within a
simple model which assumes formation of single or multiple 'hot spots`.

The junctions under study have the following structure:\ MgO//MgO$_{10nm}$
/V$_{1.5nm}$ /Cr$_{40nm}$ /Co$_{5nm}$ /Fe$_{3nm}$ /MgO$_{2nm}$ /Fe$_{5nm}$
/MgO$_{2nm}$ /Fe$_{10nm}$ /Co$_{20nm}$ /Pd$_{10nm}$ /Au$_{10nm}$. The MTJ
stacks were grown by molecular-beam epitaxy with the base pressure of
5$\times$10$^{-10}$ Torr in the presence of atomic Nitrogen. The MgO barrier
and the Fe layers were grown at RT. The Fe was annealed to 450$^{\circ}$ for
flattening. The high resolution cross-sectional TEM\ images (Fig.
1$a$)\ generally corroborate the good structural quality and homogeneity of
DMTJs. Since the TEM images are deduced from inverse Fourier transform of a
diffraction pattern integrated across the sample thickness, the amorphous
zones, if present, add a diffuse backgound to the diffraction pattern of the
single crystal zones. More details of standard double barrier samples growth
may be found in \cite{Tiusan-JCM}. The specificity of the samples studied here
is a Nitrogen doping of all the layers achieved during the growth, with a
Nitrogen concentration roughly estimated to less than 2\%. Although further
studies are needed to determine the concentration of Nitrogen inside MgO,
recent report \cite{ParkinMarch}\textbf{ }revealed\textbf{ }that MgO\ barrier
in MTJs may be doped up to 2.5\% of Nitrogen without changes in the
crystalline structure. In our samples, the structural analysis by RHEED, Auger
spectroscopy and magnetometry demonstrate that, despite the Nitrogen doping
(as evidenced from Auger, see Fig.1b), the structural and magnetic properties
are not affected. The RHEED patterns of Fe and MgO (not presented here) are
identical to those of Nitrogen-free samples. What is important for the studies
presented in this paper is that, from electrical transport point of view, the
barrier doping by Nitrogen is responsible for local 'soft' dielectric
breakdown \cite{Dimitrov} with reduced breakdown voltage. This may be expected
to keep barrier and central Fe electrode compositions nearly unchanged. After
the MBE growth of the multilayer stack, the MTJ structures were patterned to
10$\times10\mu m^{2}$ by UV lithography and Ar-ion etching, step-by-step
controlled in situ by Auger.

Figure 2$a$ shows the RT magnetization measured with magnetic field along the
Fe easy (100) axis (\textbf{EA)}. One observes well defined transitions of the
three distinct Fe layers. This is indirect indication of epitaxy and
conservation of magnetic properties (fourfold anisotropy) for Nitrogen doped
Fe. The zero bias TMR is close to 30\% (Fig.2b).

Application of the bias exceeding roughly 500 mV leads to breakdown of the
junctions, which in turn reduces zero-bias TMR down to about 4\% (see Figs.
2$c,d)$. Two other observations indicate indirectly that this breakdown
decreases the effective MgO\ barrier height most probably by transforming
locally nearly crystalline MgO\ regions near a 'hot-spot' into amorphous.
Firstly, the coercive field of the central free Fe layer remains unchanged
after breakdown (H$_{c}$=45 Oe), while coercive fields of the upper and bottom
electrodes increase substantially about 50\%. These changes are most probably
due to hardening of the Fe/Co interfaces by high current density close to the
'hot-spot'. The second indication for the possible amorphization of the MgO
barrier during the breakdown with intact middle electrode is our experimental
observation of the signature of QWS in the electron transport in some of the
broken DMTJs which we discuss below.

We have carried out detailed study of the RT resistance as a function of
\ magnetic field with three different field orientations:\ \textbf{EA}, the
hard (110) axis -\textbf{HA}, and in intermediate axis (\textbf{IA}) situated
approximately at 10 degrees from the \textbf{EA}. Here we present results
obtained at bias voltages up to 1.5V and with the steps of 25mV for those
broken DMTJs which showed reproducible signatures of the changes in the
resistance with bias (Fig.3). The zero bias TMR varies substantially along the
three mentioned directions. As expected, TMR is reduced along \textbf{IA} in
comparison with \textbf{EA} and \textbf{HA} directions, where transitions in
all three magnetic layers as indicated by arrows are reflected in the
TMR.\emph{ }We shall concentrate further on the electron transport data
obtained with field along \textbf{HA} and \textbf{IA}\ directions, where the
largest relative changes in TMR\ vs. bias were observed.

Figure 3$a$ shows typical bias dependence of the resistance $R$ for parallel
(P) and antiparallel (AP) states with magnetic field along \textbf{HA}. One
observes oscillatory behavior of R and TMR\ with a period close to 150mV in
both P and AP states. It is important to note that these oscillations, more
clearly resolved for negative bias when current flows from the upper to bottom
electrodes (Figs. 3$a,c$), have period that is in reasonable agreement with
the predictions by Wang \textit{et al.} \cite{YanWang}. Figures 3a,c mark with
arrows the majority spin QWS energies calculated within about 1V above (red)
and below (black) the Fermi energy \cite{YanWang}.

Although the absolute values of TMR measured with the field along
\textbf{IA}\ are reduced in comparison with those for the fields along\textbf{
EA} and \textbf{HA}, the relative changes of the TMR\ with bias are
substantially enhanced (see Fig. 3\emph{b,}$d$)). In order to understand this
effect we remind that the measurements of tunnel resistance with the field
along \textbf{IA}\ are usually observed \cite{X.Liu}\textbf{ }to be most
sensitive to small variations in the angle between magnetizations of the fixed
and free layers in comparison with \textbf{EA} and \textbf{HA}%
\ configurations. We suggest here that the strongest relative changes in
TMR(V) for the \textbf{IA}\ configuration could be a consequence of local
spin-torque effects which are predicted to be enhanced with intermediate
alignment of the ferromagnetic layers \cite{Theodonis}. Figures 4 $a,b$
represent 3D plots of TMR vs magnetic field and bias with magnetic fields
applied along \textbf{HA} and \textbf{IA}\thinspace\ respectively. Dependence
of TMR\ on bias is observed to be more asymmetric with field along
\textbf{IA}\ (Figs. 4 $a,b$), which is in agreement with possible influence of
local spin torque effects in the breakdown regions.

Let us now discuss physical mechanisms which could be behind the main
experimental findings. Before breakdown, the current is roughly uniform across
the junction area and weak interface disorder might introduce decoherence
suppressing the effects due to QWS. The 'soft' breakdown of the doped MgO may
create defects and local amorphization which locally reduce the MgO barriers
leading to 'hot spots' which connect Fe and Fe/Co leads with the central Fe
layer. With 'hot spots' of sufficiently small lateral dimensions, electrons
tunneling to the central layer can sample well defined structure of QWS due to
lack of decoherence. The remaining part of the macroscopic DMTJ provides then
some averaged featureless background signal. The feasibility of the above
scenario is qualitatively supported by the good correspondence of the observed
periodic variations in R vs.V and the theory taking into account QWS formed
within 4.6 nm thick central Fe electrode \cite{YanWang} (see Figure 3a,c).

In order to describe the observed features, we consider a DMTJ (Fig.~4c,d)
with barriers including a number of 'hot spots'. The average conductance of
the structure is \cite{Lifshitz} $\sigma\simeq\sum_{i}\int p_{i}(\Gamma
_{i})\,\sigma_{i}(\Gamma_{i})\,d\Gamma_{i},$where $\sigma_{i}(\Gamma_{i})$ is
the conductance due to a single $i$-th spot, $p_{i}(\Gamma_{i})$ is the
probability of realization of a certain configuration of the $i$-th spot, and
$\Gamma_{i}$ is a set of parameters characterizing this configuration. Let us
assume that a particular 'hot spot' is characterized by its lateral dimension
$a$. The current through the spot can be then calculated as
\begin{align}
I  &  =\frac{e\hbar}{m}\int_{\varepsilon_{F}-eV}^{\varepsilon_{F}}%
d\varepsilon\int\frac{d^{3}\mathbf{k}_{1}}{(2\pi)^{3}}\;\delta\left(
\frac{\hbar^{2}(k_{1}^{2}+k_{1l}^{2})}{2m}-\varepsilon\right)
\nonumber\label{2}\\
&  \times\sum_{k_{2}<k_{2m}}w_{\mathbf{k_{1}k_{2}}}(a)\,T_{k_{1}k_{2}}%
T_{k_{2}k_{3}}k_{3},
\end{align}
where $k_{1},\mathbf{k}_{1l}$ are the normal and lateral wavevector components
of the incoming wave (layer 1), $k_{2}$ and $k_{3}$ are the normal components
of the wavevectors in layers 2 and 3, respectively. Here we assumed that the
in-plane component of $\mathbf{k}$ is conserved for tunneling from layer 2 to
3, whereas for tunneling through the spot there is no conservation of the
in-plane component due to broken translational symmetry, and the scattering
can be described by an angle distribution function $w_{\mathbf{k_{1}k_{2}}%
}(a)$. In our calculations we use the approximation $\label{3}w_{\mathbf{k_{1}%
k_{2}}}(a)\simeq a^{2}e^{-ak_{2l}}$, which means that scattering to the state
with large in-plane component $k_{2l}$ is effectively suppressed. Equation (1)
includes transmission probabilities $T_{k_{1}k_{2}}$ and $T_{k_{2}k_{3}}$ for
tunneling from the layer 1 to 2, and from 2 to 3, respectively. The sum over
$k_{2}$ runs over discrete values satisfying the quantization condition
$k_{2}L=n\pi$. It should be emphasized that this condition is related only to
thickness $L$ of the layer 2 and is the same for any other 'hot spot'.

Calculating the integral over $k_{1l}$ we find
\begin{equation}
I=\frac{ea^{2}}{\hbar}\int_{\varepsilon_{F}-eV}^{\varepsilon_{F}}%
d\varepsilon\int_{0}^{k_{1m}}dk_{1}\sum_{k_{2}<k_{2m}}e^{-ak_{2l}}%
T_{k_{1}k_{2}}T_{k_{2}k_{3}}k_{3}.
\end{equation}
The conductance $I/V$ as a function of bias $V$ for a single spot is presented
in Fig.~4e for different values of $a$. We assumed $L=4.6$~nm, the barrier
width $L_{B}=2.4$~nm, $\varepsilon_{F}=0.9$~eV, and the barrier height
$U_{B}=\varepsilon_{F}+3.8$~eV. As we see, the oscillation peaks related to
the level quantization in the layer 2 are more pronounced for wide spots, and
they are effectively damped for small $a$. This is because the small spot
enables tunneling with nonconserved in-plane component of the wave vector.

Taking into account tunneling from many different spots, we obtain
qualitatively the same picture corresponding to a mean value of $a$, and
proportional to the number of spots. The total conductance of the structure
includes a constant non-oscillating part, $\sigma_{0}$, related to the
tunneling without spots. In Fig.~4f we present resistance calculated as
$R=(\sigma_{0}+N_{i}\sigma_{i}(\overline{a}))^{-1}$, where $N_{i}$ is the
number of spots and $\overline{a}$ is the mean value of $a$. We note that the
variation of barrier heights at the spot does not affect the position of peaks
and does not change the shape of peaks, changing only the amplitude. Thus,
averaging over randomly distributed $a$ and barrier heights $U_{B}$ gives a
picture like for a single spot with some mean values of $a$ and $U_{B}$.

The applicability of the above model requires a number of conditions. Although
location of the oscillation peaks does not depend on the spot dimension ($a$),
the 'hot spot' should not exceed average dimensions over which central
electrode is atomically flat. Also, as seen from Fig.4e,f, decreasing the
parameter $a$ makes oscillations less visible (damped). Therefore we expect
the hot spots to be roughly of \emph{nm} lateral size. Secondly, if magnetic
moment of one or two layers is reversed (AP alignment,) the resistance becomes
very large because the minority $\Delta_{1}$ band will be displaced well above
the Fermi energy. However realistically speaking the hot spot region of the
MgO barrier may be not fully epitaxial either due to structural defects and/or
Nitrogen doping. This can strongly reduce spin filtering by the $\Delta_{1}$
band in the AP alignment, substantially suppressing TMR in real DMTJs with
'hot spots'.

In conclusion, we have presented evidence for local resonant tunneling through
quantum well states in the middle continuous free layer of a double magnetic
tunnel junction. The oscillations have been observed at room temperature in
both parallel and antiparallel magnetic configurations and for both bias
polarizations. Owing to specific features of the breakdown junctions, we were
able to observe quantum well states in continuous magnetic layers. To observe
similar effects in the Nitrogen free DMTJ, the junction area should be smaller
than the size of terraces of the Fe central layer. Understanding of electron
transport in magnetic tunnel junctions with defects and 'hot spots' is of
great importance both from fundamental and applied points of view. Recent
reports link spin torque oscillations with record low bandwidth to the
presence of defects and hot spots inside the MgO barrier of MTJs
\cite{HoussameddinePRL}.

We thank S.~Parkin and J.~Martinek for stimulating discussions, F.Greullet for
help with experiment and E.Snoeck for taking TEM images. The work was
supported by Spanish MICINN (MAT2009-10139, MAT2006-28183-E, CSD2007-0010,
Integrated Action project FR2009-0010) and CAM (P2009/MAT-1726), by the FCT
Grant PTDC/FIS/70843/2006 in Portugal and by Polish Ministry of Science and
Higher Education as a research project in years 2007 -- 2010.

\section{FIGURE\ CAPTIONS}

\bigskip%

\begin{center}
\begin{figure}[h]
    \begin{center}
    \includegraphics[width=10cm]{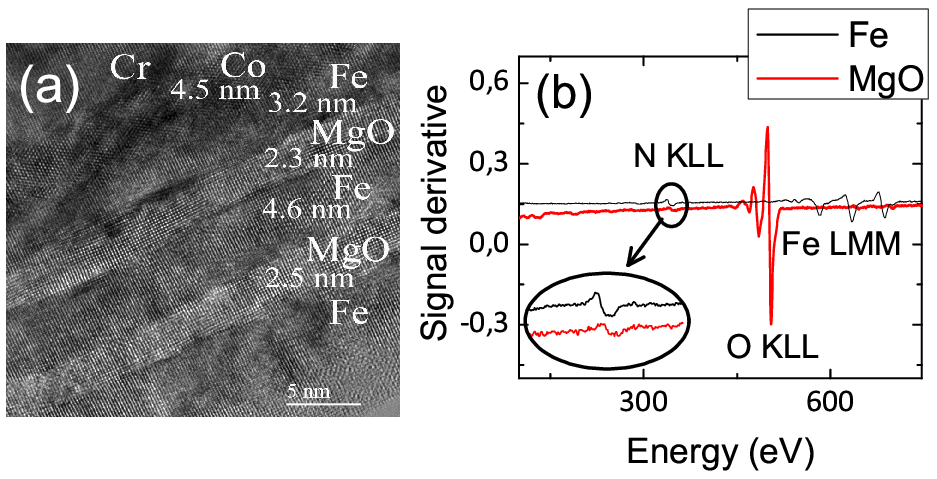}
     \caption{(a) Cross-sectional TEM image of the DMTJ. (b) Auger data which shows presence of Nitrogen doping in the Fe and MgO
     layers of DMTJs}.
\end{center}
\end{figure}
\end{center}

\begin{center}
\begin{figure}[h]
    \begin{center}
    \includegraphics[width=10cm]{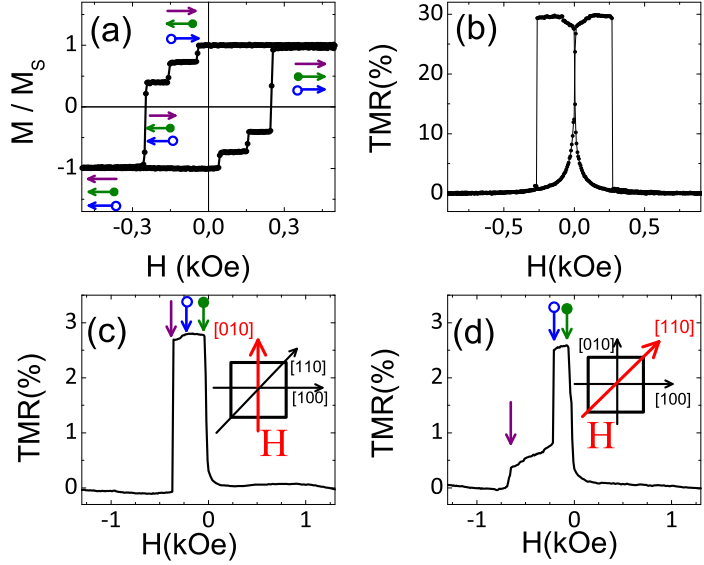}
     \caption{(a) Magnetization curve for unpatterned DMTJ. Arrows
indicate magnetization configurations of top (black), middle (green) and
bottom (red)\ layers. (b)\ Typical zero bias TMR of the DMTJ without hot spot
measured along the \textbf{EA}. Figures (c) and (d) show correspondingly
zero-bias TMR for field along the\textbf{ EA}, and the \textbf{HA} in
junctions with hot spots. The vertical arrows remark coercive fields.}.
\end{center}
\end{figure}
\end{center}

\begin{center}
\begin{figure}[h]
    \begin{center}
    \includegraphics[width=10cm]{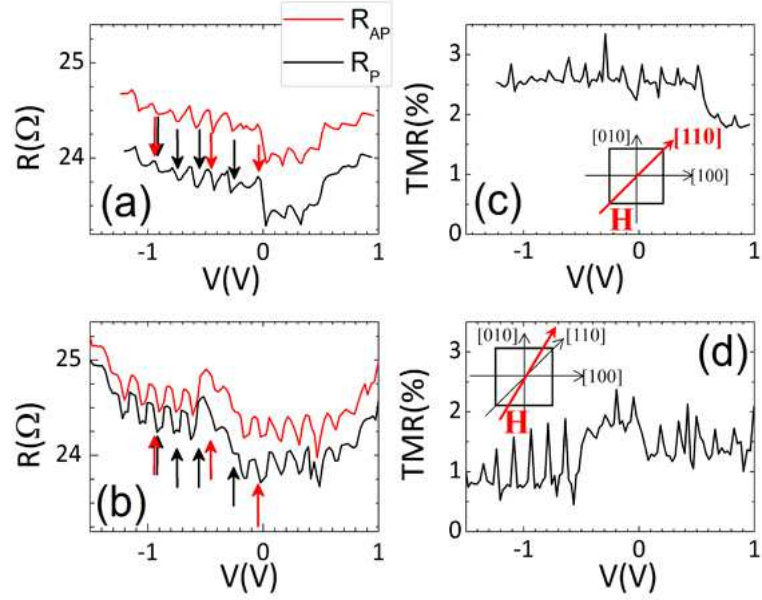}
     \caption{Resistance vs. bias for P (black line) and AP (red line)
states with magnetic field along the \textbf{HA} (a)\ and \textbf{IA} (b).
Here we assign as AP state the one just after inversion of the central Fe
layer (green arrow in Fig.2). The red and black arrows indicate predictions by
Wang \textit{et al.} \cite{YanWang} for the resonant tunneling in the parallel
state with QWS above (red) and below (black) the Fermi level. Parts (c,d) show
TMR vs. bias for magnetic field along the \textbf{HA} and \textbf{IA}.}.
\end{center}
\end{figure}
\end{center}

\begin{center}
\begin{figure}[h]
    \begin{center}
    \includegraphics[width=10cm]{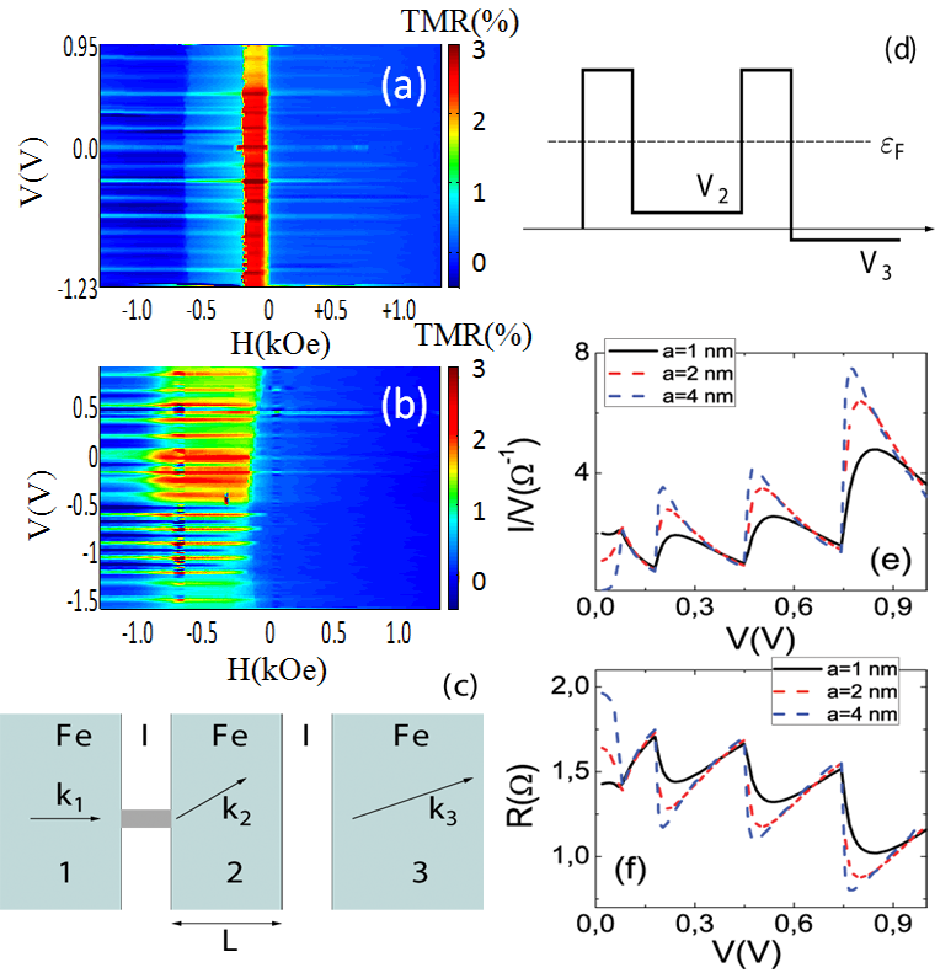}
     \caption{3-D-plot with \ magnetic field along $x$, bias voltage
\ along $y$ and TMR\ along $z$ directions. Part (a) corresponds to magnetic
field along the \textbf{HA\ }and part (b) along \textbf{IA.} (c) Schematic
presentation of the model with a single spot and (d) the corresponding energy
profile. (e)\ Calculated conductance ($I/V$) for DMTJ with a single spot in P
state for different spot dimensions. (f)\ R vs. V for DMTJ with multiple spots
in the P state for different average spot dimensions.
}.
\end{center}
\end{figure}
\end{center}


\begin{thebibliography}{99}                                                                                               %


\bibitem {Butler}W. Butler, et. al, \textit{Phys. Rev. B} \textbf{63}, 054416 (2001).

\bibitem {Mathon}J. Mathon, et. al, \textit{Phys. Rev. B} \textbf{63},
220403(R) (2001).

\bibitem {Bowen}M. Bowen, et. al, \textit{Appl.Phys.Lett}. \textbf{79}, 1665 (2001).

\bibitem {Faure-Vincent}J. Faure-Vincent, et. al, \textit{Appl.Phys.Lett}.
\textbf{82}, 4507 (2004).

\bibitem {Parkin}S. Parkin, et. al, \textit{Nat.Mater} \textbf{3}, 862 (2004).

\bibitem {Yuasa}S. Yuasa, et. al, \textit{Nat.Mater} \textbf{3}, 868 (2004).

\bibitem {GuerreroAPL}R. Guerrero, et. al, \textit{Appl.Phys.Lett}.
\textbf{91}, 132504 (2007).

\bibitem {Moodera}J. Moodera, et. al, \textit{Phys. Rev. Lett }\textbf{74},
3273 (1995).

\bibitem {LeClair}P. LeClair, et. al, \textit{Phys. Rev. Lett} \textbf{84},
2933 (2000).

\bibitem {YuasaScience}S. Yuasa, et. al, \textit{Science} \textbf{297}, 234 (2002).

\bibitem {Greullet}F. Greullet, et. al, \textit{Phys. Rev. Lett} \textbf{99},
187202 (2007).

\bibitem {Barnas}J. Barnas, et. al, \textit{Phys. Rev. Lett }\textbf{80}, 1058 (1998).

\bibitem {Petukhov}A. Petukhov, et. al, \textit{Phys. Rev. Lett} \textbf{89},
107205 (2002).

\bibitem {Kalitsov}A. Kalitsov, et. al, \textit{Phys. Rev. Lett} \textbf{93},
046603 (2004).

\bibitem {VedyaevA}A. Vedyaev, et. al., \textit{Phys. Lett. A} \textbf{355},
243 (2006).

\bibitem {Theodonis}I. Theodonis, et. al, \textit{Phys.Rev.B} \textbf{76},
224406 (2007).

\bibitem {Watanabe}M. Watanabe, et. al, \textit{Appl.Phys.Lett.} \textbf{92},
082506 (2008).

\bibitem {PeraltaRamos}J. Peralta-Ramos, et. al, \textit{Phys. Rev. B
}\textbf{78}, 024430 (2008).

\bibitem {G.Feng}G. Feng, et. al, \textit{J. Appl. Phys.} \textbf{105}, 07C926 (2009).

\bibitem {Tiusan-JCM}C. Tiusan, et. al, \textit{J.Phys.: Cond. Mat.}
\textbf{19}, 165201 (2007).

\bibitem {Coey}G. Feng, et. al, \textit{Appl.Phys.Lett. }\textbf{89}, 162501 (2006).

\bibitem {Iovan}A. Iovan, et. al, \textit{NANO LETTERS} \textbf{8}, 805 (2008).

\bibitem {Nozaki-PRL}T. Nozaki, et. al, \textit{Phys.Rev.Lett.} \textbf{96},
027208 (2006).

\bibitem {YanWang}Y. Wang, et. al, \textit{Phys. Rev. Lett} \textbf{97},
087210 (2006).

\bibitem {ParkinMarch}S. Parkin, \textit{APS2009 March Meeting},
\textbf{Z32}.00001 (2009).

\bibitem {Dimitrov}D. Dimitrov, et. al, \textit{Appl.Phys.Lett}. \textbf{94},
123110 (2009).

\bibitem {X.Liu}X. Liu, et. al, \textit{J. Appl. Phys}. \textbf{92, }8 (2002).

\bibitem {Lifshitz}I. Lifshitz, et. al, \textit{Sov. Phys. JETP} \textbf{50},
499 (1979).

\bibitem {HoussameddinePRL}D. Houssameddine, et.al, \textit{PRL} \textbf{102},
257202 (2009).
\end{thebibliography}
\end{document}